The Dynamic Replicon: adapting to a changing cellular environment

Submitted July 8[th] 2008 to Frontiers in Bioscience

John Herrick
3, rue des Jeûneurs
Paris 75002
jhenryherrick@yahoo.fr

Abstract: Eukaryotic cells are often exposed to fluctuations in growth conditions as well as endogenous and exogenous stress-related agents. In addition, during development global patterns of gene transcription change dramatically, and these changes are associated with altered patterns of DNA replication. In metazoan embryos, for example, transcription is repressed globally and any sequence in the genome can serve as a site for the start of DNA synthesis. As transcription is activated and a G1 phase imposed, the pattern of replication adapts to these changes by restricting the sites where DNA synthesis begins. Recent evidence indicates that each unit of replication, or replicon, is specified by two or more potential replication origins, but only one is selected to initiate replication of the replicon. How the cell distinguishes between potential origins, and how it selects a given origin of replication remain unclear. This raises important questions concerning the nature and definition of the eukaryotic replicon. In the following we will review emerging evidence concerning the mechanisms involved in regulating replication origins during both the normal and perturbed eukaryotic cell cycle.

Introduction

The replicon hypothesis, formulated in 1963, proposed that the replication of DNA consists of two principal features: a *cis*-acting factor present on the molecule, called the replicator, which serves as a genetically defined site where DNA synthesis begins; and a *trans*-acting factor, called the initiator, which performs a regulatory function in activating duplication of the replicon (1). According to this simple hypothesis, the replicon consists of a single molecule of DNA that is replicated bi-directionally from a discrete site called a replication origin. The basic features of this proposal are now well established in bacteria (2). The bacterial genome contains one or more separate replicons that initiate DNA synthesis from a single site (or a few restricted sites), and are duplicated bi-directionally by a pair of moving replication forks. The activation of DNA synthesis is regulated *in trans* by a positively acting factor, the initator protein, that signals the start of DNA synthesis when it binds to the replication origin.

While the hypothesis has been applied successfully to simple organisms such as viruses and bacteria, more complicated organisms have resisted attempts to explain genome duplication on its basis (3). Eukaryotes have large complex genomes that rely on multiple origins of replication to assure complete duplication of their full genetic complement (4). In the yeast *Saccharomyces cerevisiae*, a consensus DNA sequence that permits duplication of DNA, a replicator, has been identified. This sequence, called an autonomously replicating sequence (ARS), confers on otherwise non-autonomously replicating DNA the ability to undergo replication (5, 6)

Replication in eukaryotes is initiated during a complex process that begins with the binding of a *trans* acting complex, called the origin recognition complex (ORC) (7). While ORC homologues have been identified in all eukaryotes examined so far, specific replicators in eukaryotes other than *S cerevisiae* have remained elusive. For example, only a dozen or so ARSs have been identified to date in metazoans (8). The failure to identify or clone replicators in eukaryotes is not due to technical difficulties alone, but rather to the lack of a genetically defined sequence that determines the start site of DNA synthesis (9). This observation indicates a radically different mode of DNA replication in eukaryotes compared to prokaryotes.

Studies conducted in the 1960s and 1970s established the eukaryotic paradigm of DNA replication (10, 11). These studies relied in large part on a technique called DNA fibre autoradiography that permitted the direct labeling of replication forks, and, by implication, the direct visualization of replication origins in eukaryotes. Although origins of replication could not be mapped to their genetic sites using this technique, the paradigm established that the eukaryotic genome is organized in multiple, tandem replicons each containing a single replication origin and each replicated bi-directionally, and in some cases unidirectionally, by steadily advancing replication forks. These studies were notable for establishing that 4 to 5 tandem replicons correspond to a higher order of organization, termed replicon clusters, inside discrete nuclear sites called replication foci (12, 13). Replicon clusters are revealed by the more or less synchronous replication of adjacent replicons. It remains unclear, however, if the synchronous activation of replicons is merely a consequence of the architechtural organization of replicons inside replication foci, or if it is of regulatory importance, and hence necessary for effective genome duplication (14, 15).

Recent developments relying on DNA fibre fluorography have enabled researchers to quantitatively examine the kinetics of genome duplication during S phase of the cell cycle (16, 17, 18, 19, 20, 21). Early results on the *Xenopus laevis in vitro* embryonic DNA replication system revealed that origins of replication in that system are closely spaced at intervals of 5 to 15 Kb and asynchronously activated throughout S phase (17, 18, 19). The same studies also revealed that two distinct replication regimes govern duplication of the genome, and that replication origins are activated stochastically as S phase advances (17, 18)

The observation that replication origins dispersed throughout the genome are stochastically activated has recently been extended to two other eukaryotic systems, *Schizosaccharomyces pombe* and *Saccharomyces cerevisiae* (22, 23). The stochastic activation of replication origins in eukaryotes reflects the fact that replication origins in these systems exist in excess of the number required to complete genome duplication. In *S cerevisiae*, for example, over 10,000 consensus ARS sites have been identified, yet only 400 are used during a given S phase (24). Several other observations have shown that these redundant origins are activated in response to DNA damage and other impediments to replication fork movement (25, 26, 27). Together, these observations indicate that eukaryotic replicons are considerably more complex than those found in prokaryotes, and are able to spontaneously adapt to unexpected events such as the interruption of a replication fork or the misfiring of a replication origin as S phase advances. This review will examine origin usage and flexibility in eukaryotes during both normal and perturbed cell division cycles. Particular emphasis will be placed on the regulation of dormant replication origins.

The hierarchical organization of replication origins in the genome

Early studies on DNA replication in eukaryotes indicated that new sites of replication are recruited when DNA synthesis is inhibited (25, 26). Blocking entry into S phase, for example, resulted in a marked reduction in replicon size, and the extent to which replicon size decreased depended on the amount of time the cells were blocked at the G1/S phase transition. These studies revealed that the minimal replication origin spacing was 12 kb (compared to an average of 220 kb under normal conditions) in a Chinese hamster (CHO) cell line (25, 26).

Later studies confirmed these findings in CHO cells and demonstrated that initiation of DNA replication occurred at the same sites when replication forks were inhibited in two related cell lines. These studies examined the activation of a previously characterized origin of replication near the AMPD2 gene, which is amplified up to 100 fold in this cell line (28). Hydroxyurea (HU), which inactivates ribonucleotide reductase and blocks dNTP synthesis (29), was used to inhibit replication forks; and origin densities were found to increase approximately two fold. At the same time, replication fork rates decreased proportionally. This and other early studies indicate that the cell relies on a set of backup, or dormant, origins that are activated in response to replication anomalies. The repeated use of the same dormant origins in the HU studies suggested that dormant origins correspond to well-defined sites in the genome rather than to diffuse regions where DNA synthesis is randomly activated.

Two other studies also employing hybridization techniques to study origin usage revealed that individual replicons are specified by multiple potential origins, and that an active origin in a replicon is selected stochastically without predefined timing preferences (30, 31). Experiments on human primary keratinocytes employed a novel hybridization technique termed Genomic Morse Code to map origins in a 1.5 Mb region of chromosome 14q11.2 (31). These experiments revealed a hierarchical organization of potential replication origins within individual replicons. This hierarchy was found to consist of 3 basic levels: 1) potential origins of replication are located anywhere within initiation zones of 2.6 to 21.6 Kb; 2) replicons correspond to 2 to 4 initiation zones, but only one potential origin in a zone specifies a replicon; and 3) replicons are grouped together in clusters; however, the firing of replication origins is not temporally correlated, and origins in a cluster fire independently of each other.

The other study investigated origin usage in the mouse *igh* locus during development of B cells in order to understand the relationship between the regulation of replication origins and other developmental events (30). These studies revealed that origins of replication are activated in clusters, and major changes in the pattern of origin activation take place during development. The observed changes were able to account for the replication timing of sub-regions of the locus, and therefore excluded changes in replication fork rates in regulating replication timing. The studies also revealed that replication origins were activated on average every 100 Kb, but potential origins are spaced every 20 Kb: or approximately 3 to 4 potential initiation sites per replicon. Consequently, a large number of origins are available for initiation; but the frequency, or efficiency, at any particular origin was low, indicating the absence of a predominant initiation site. It was suggested that the small number of potential origins used during a particular S phase might reflect the existence of a checkpoint that limits the total number of active replication forks at any given time within the cluster.

The observations on the spacing between potential origins agree closely with those found in the human keratinocytes, and are also in agreement with earlier reports on origin spacing in *Xenopus laevis* embryos. In keratinocytes, the size of initiation zones varies significantly and the distance between the centers of adjacent zoness was found to be 40 Kb +/- 20 Kb (Figure 1). The gaps between the zones, however, where no initiation events took place, were significantly more regular, suggesting a non-random organization of zones that are refractory to initiation. The average size of the gaps, which reflects the distance between potential replication origins in initiation zones, corresponded to an average value of approximately 25 Kb, in agreement with the observations on the *igh* locus. This indicates that replication is organized into alternating zones where initiation is either completely prohibited or otherwise permitted anywhere inside the zone. One possible explanation for this periodic pattern of origin spacing is suggested by the observation that in both the *igh* and keratinocyte systems a marked preference for initiation occurs in intergenic regions (Figure 1). Transcriptional activity therefore potentially influences the locations where origins fire, although there appear to be many exceptions.

The role of the checkpoint in regulating replicon size

The molecular basis for the distinction between normal, or preferential, origins and dormant origins remains at present to be elucidated. It has been noted for some time, however, that the MCM proteins, which are essential for origin activation, exist in a 20 to 40 fold excess over ORC complexes bound to chromatin (32, 33). Recently, it has been shown that the excess MCM proteins license dormant origins in replicating *X laevis* extracts and human cells exposed to a variety of replication fork inhibitors (34, 35). It was further shown that dormant origins are essential for the cell to survive replicative stress, suggesting that dormant replication origins represent another component in the cell's repertoire of DNA damage response (DDR) elements (35). Although dormant origins appear to be specified independently of the ORC complex, other studies suggest that ORC itself exists in excess of what is required for genome duplication (36). This observation might provide one explanation for the molecular distinction between normal and dormant origins. Accordingly, a replicon corresponds to one or more ORC specified preferential origins, which are "backed up" by a subset of ORC independent, MCM specified dormant origins (Figure 2).

Two possible mechanisms of dormant origin activation, among others, can be envisioned: 1) dormant origins are activated passively when replication is delayed through the region (35); and 2) dormant origins are activated when origin interference mechanisms are relaxed following prolonged replicative stress (15, 37, 38, 39, 40, 41, 42 and references therein). Although these mechanisms differ they are not mutually exclusive and the cell likely employs both. The former mechanism was proposed based on the observation that origin densities increase when replication fork movement is impeded, and hence the probability of activating a dormant origin increases as a function of the time the chromatin remains unreplicated (25, 26, 43, 44). According to this interpretation, dormant origins fire passively in the presence of active checkpoint surveillance (35). Hence, origins fire with increasing probability when either S phase or fork movement is delayed.

The latter mechanism was proposed based on observations that the local activation of the checkpoint and its down-regulation in replicating regions of the genome are coupled via the degradation of Chk1 and possibly claspin (41, 42). According to this interpretation, dormant origins fire after checkpoint functions have been either attenuated or experimentally abrogated (35). A third interpretation incorporates aspects of both mechanisms, namely: dormant origins fire passively in a checkpoint independent manner that up-regulates the checkpoint (1 to 4h of inhibition); and prolonged fork blockage and DNA damage stimulates checkpoint down-regulation and the firing of additional dormant origins (4 to 8h of inhibition). Accordingly, dormant origins are fired in two separate phases prior to checkpoint activation (passively) and after checkpoint attenuation (actively). What is the evidence in support of this proposal?

During a normal S-phase, the experiments on keratinocytes revealed that origin interference occurs as a result of the replication through and passive inactivation of nearby potential origins (31). The passive mechanism of origin intereference therefore appears to determine replicon size during a normal S phase when checkpoint surveillance is operational but the checkpoint itself is inactive. This is consistent with the proposal that dormant origin activation also occurs passively when a sufficient amount of time elapses before a region containing blocked forks can be replicated. The later firing of weak, or inefficient, origins agrees with the observation that origin densities increase as a function of the time DNA replication is inhibited (25, 26). Moreover, a passive mechanism can account for the observed correlation between origin densities and fork rates, because slower forks will be proportionally compensated by higher origin densities. This mechanism implies that the checkpoint is "by-passed" when forks stall. Fork rates and origin densities are therefore independently regulated, but nevertheless coordinated through the time dependent nature of dormant origin activation.

Dormant origin activation as a component of checkpoint recovery

A more direct mechanism regulating dormant origin firing is based on the observations that Chk1 actively imposes origin interference in its surveillance mode during normal replication (37, 38, 39, 40). Evidence for an active mechanism involving the checkpoint mediator Chk1 has been reported in a number of different studies (33, 34, 35, 36). These experiments revealed that abrogating the checkpoint results in shorter inter-origin distances and a correspondingly reduced fork rate. Concomitantly, levels of chromatin bound Cdc45 increase up to 20 fold in the presence of aphidicolin and the ATR/Chk1 pathway inhibitor caffeine (34). The increase in Cdc45 loading was even more pronounced in the presence of actinomycin D, a DNA primase inhibitor (35). The latter observation can be explained if actinomycin D prevents checkpoint activation in addition to its role in blocking replication forks, because robust checkpoint activation depends on the synthesis of DNA primers on single strand DNA (45). Under these conditions, the checkpoint will be "short-circuited" and dormant origins will subsequently fire. Together, these observations suggest that Chk1 functionally modulates the levels of chromatin associated Cdc45, and attenuation of Chk1 function is thus associated with dormant origin firing.

Additional evidence that dormant origins are actively regulated comes from two separate sets of experiments. The first set of experiments revealed that dormant origins fire when translesion DNA polymerases are over-expressed and replication fork rates simultaneously decrease (46). Under these conditions, the checkpoint is not activated although fork movement is retarded. The other set of experiments showed that under conditions of replicative stress, the translesion polymerases, which are regulated by the Rad6/Rad18 ubiquitin-conjugating complex, replicate through DNA lesions, and consequently attenuate, or relax, the checkpoint (47, 48, 49). In these experiments, mouse embryo fibroblasts (MEF) were treated for 2h with either BPDE treatment, a DNA alkylating agent, or exposed to UV. These two genotoxic agents reduced DNA synthesis by 40% in $Rad18^{+/+}$ cells. However, 4 h post-BPDE treatment, rates of DNA synthesis recovered to control levels. In $Rad18^{-/-}$ cells, BPDE inhibited DNA synthesis with kinetics similar to those of WT MEFs, but DNA synthesis failed to recover to control levels within the time frame of the experiment. This effect did not occur when cells were treated with either HU or ionizing radiation (IR). These experiments, therefore, are not consistent with the passive mechanism of dormant origin firing, and they suggest that replication recovery following DNA damage depends on active checkpoint attenuation mediated by Rad18/Rad6 ubiquitination pathway. Whether or not recovery depends on replication restart at stalled replication forks or dormant origin activation, or both, remains to be investigated.

Other experiments in yeast led to similar findings. In *S cerevisiae*, the cullin Rtt101p promotes replication fork progression through damaged DNA. It was shown that in *rtt101Δ* cells, unreplicated DNA persists, and the cells accumulate spontaneous DNA damage and exhibit a $G_2$/M delay (50). Under these conditions, the Chk1 functional orthologue Rad53 is activated (51), and late replication origin firing is subsequently blocked (52). Although these effects were not attributed to a defect in dormant origin activation, they are consistent with prolonged checkpoint up-regulation, and hence persistent suppression of dormant origin firing in response to blocked replication fork movement. Together, these experiments lend additional, albeit indirect, support to the proposal that dormant origin activation coincides with Chk1/Rad53 degradation and checkpoint down-regulation.

The kinetics of Chk1 degradation in each of these experiments also supports the proposal that dormant origin activation occurs in response to proteasome mediated checkpoint attenuation. In the experiments showing checkpoint independent activation of dormant origins, Chk1 was activated within minutes of exposure to the topoisomerase inhibitor CPT, but degradation did not occur until 4h to 8h after drug treatment (53). In contrast, when hydroxyurea (HU) is used to inhibit fork movement, 4 hours of exposure to the drug resulted in a decrease in the distance between forks by nearly two fold (35), but did not reveal a clear degradation of activated Chk1 at 4h. The discrepancy might be due either to the different DNA damage pathways involved, or it might reflect a more localized degradation of Chk1 that is not detectable at 4h of genotoxic stress. Therefore, the connection between dormant origin firing and a more global Chk1 down-regulation after 4 h remains to be verified.

Although the limited time frame of these experiments did not permit detection of a more global down-regulation of Chk1, other experiments carried out in parallel did reveal that knocking down Chk1 resulted in a very similar increase in fork density (35). This suggests that Chk1 down-regulation is involved in dormant origin firing. The results of this experiment therfore agree with the earlier experiments that Chk1 activity suppresses dormant origin firing. Taken together, the above results are consistent with the observations that prolonged levels of genotoxic stress result in: 1) coupled activation and degradation Chk1; and 2) higher origin densities that coincide with checkpoint down-regulation and enhanced chromatin loading of Cdc45. The observed time dependent activation of dormant origins can therefore be explained by the kinetics of Chk1 degradation in addition to the proposed "over-riding" of Chk1 mediated surveillance by a passive mechanism involving the time-dependent accumulation of an initiation factor.

The role of oncogene over-expression during dormant origin activation

Chk1 over-expression has been shown to result in reduced levels of Cdc45 bound chromatin as a consequence of inhibiting Cdc7/Dbf4 (54, 55). Conversely, inhibiting Cdk2 has been coincides with an excess loading of MCM proteins (56, 57), an observation that is consistent with earlier findings that high levels of cdk2-cyclin E within nuclei prevent MCM proteins from re-associating with chromatin after replication (58). When Cdk2 levels were experimentally reduced, re-replication occurred and Chk1 was activated, suggesting that the enhanced loading of MCM proteins coincided with unscheduled origin firing. Chk1 degradation, however, was not investigated during these studies, so it remains unclear if its coupled activation and degradation plays a role in dormant origin activation under these conditions.

A recent paper reports that MYCN over-production stimulates MCM protein expression (59), and that their levels are correlated in a variety of cancer cell lines. Myc overexpression also results in elevated levels of Cdk2-cyclinE and accelerated progression through G1 (60, 61). These observations are consistent with the finding that MYC overproduction induces unscheduled replication, gene amplification and DNA fragmentation, and suggest that this occurs via MCM gene regulation and/or regulation of CDK2 activity (62, 63, 64). CMYC depletion, on the other hand, causes a significant decrease in the number of active replicons, suggesting that MYC plays a role in specifying and/or activating replication origins (65, 66). Thus, CMYC induced genome instability is determined in G1 but occurs during S phase, possibly as a result of excessive origin licensing (67, 68).

In addition to MYC, dormant origin activation occurs when the Ras oncogene is over-expressed, resulting in a phenomenon termed hyper-replication (69). Hyper-replication has been proposed to act as a mechanism to amplify the DNA damage response (DDR) and induce replicative senescence (70). Failure to engage the DDR results in genomic instability and cellular transformation. Hyper-replication is associated with the production of reactive oxygen species (ROS), which implicates elevated levels of mitochondrial activity as a component of the DDR and suggests a mechanistic link between HR, the DDR and apoptosis. The proposal that ROS stimulates the DDR, perhaps by signalling HR, and induces cellular senescence is consistent with other findings that inducing mitochondrial respiration increases oxidative stress and, paradoxically, extends life-span (71). The extension of life span as a result of ROS production was explained by the ROS dependent induction of the DDR, and hence enhanced protection of the genome against replicative stress, a phenomenon referred to as "hormesis".

These results suggest that oncogene over-expression can likewise over-ride, or by-pass, the checkpoint and induce checkpoint independent HR. Accordingly, three different modes of checkpoint surveillance can be defined (15, see figure 5): 1) origin activation at G1/S phase induces a surveillance mode that regulates fork densities and origin spacing; 2) low levels of genotoxic stress or oncogene over-expression moderately up-regulates the checkpoint, which stabilizes replication forks and facilitates DNA repair and replication restart; and 3) chronic levels of stress cause a down-regulation of Chk1, which coincides with a transient period of hyper-replication followed by replication fork collapse, a strong DDR (Chk2 induction) and irreversible S phase arrest (OIS or apoptosis). Hence, HR involves two phases both of which are checkpoint independent: an early, and more local, ROS dependent phase that induces the DDR, and a later, more global phase that depends on Chk1 degradation, which potentially induces apoptosis under conditions of prolonged stress and incomplete replication (72).

In contrast, during a normal unstressed S phase, initiation factors such as Cdc45/Cdc6/Cdt1 are inactivated as S phase nears completion, resulting in a rapid decline in imitation frequency. Consequently at the S/G2 transition, elongation terminates in the absence of initiation, which prevents re-replication during G2/M. The decrease in initiation frequency that is expected is presumably due to proteosomal degradation of essential initiation factors (68), as occurs during the M to G1 transition. A rapid decrease in initiation frequency at the end of S phase is consistent with APC/C regulation of origin licensing (Figure 2).

The role of dormant origin activation in regulating fragile site expression and genome stability

The role of chromosomal instability (CIN) in promoting cellular transformation has long been a subject of debate concerning whether it is a cause or a consequence of carcinogenesis (73, 74). Early models of gene amplification proposed that hyper-replication caused DNA breakage associated with gene amplification and subsequent carcinogenesis (75, 76, 77, 78). Although DNA replication based models of gene amplification were later abandoned in favour of recombination based models (79, 80), other models implicate breakage of replication bubbles in the formation of extra-chromosomally amplified double minute chromosomes (Figure 3; 81, 82, 83). More recent studies also implicate hyper-replication followed by replisome collision as an initiating step in DNA fragmentation and recombinational amplification (84). A mentioned above, MYC overproduction has been shown to induce re-replication and gene amplification (85), and it will be interesting, therefore, to determine its exact role, if any, in regulating excessive licencing and dormant origin activation.

Based on observations that replication intermediates accumulate in difficult to replicate regions of the genome (86), it has recently been proposed that the induction fragile sites by DNA synthesis inhibitors such as aphidicolin might occur as a result of dormant origin activation (15). This proposal is consistent with the replisome collision model of DNA fragmentation (84), since an overall reduction in origin spacing in response to slowing replication forks would increase the likelihood of a collision between stalled and moving replication forks. Consistent with such a proposal, depletion of Chk1 but not Chk2, was found to induce chromosome instability and breaks at fragile sites (87), suggesting that dormant origin activation might occur at a higher frequency at fragile sites than elsewhere in the genome when Chk1 is depleted and dormant origins fire (Figure 4).

A recent report also implicates the Fanconi anemia pathway in the activation of dormant origins when replication forks stall (88). These studies showed that replication restart in FANCD+ cells occurs via the activation of new replication origins, whereas replication in FANCD- cells resumed from the stalled forks themselves. In agreement with the general principle of activating new origins when replication forks stall, it was proposed that the FA pathway arrests replication forks during genotoxic stress and promotes the firing of new origins after removal of the replication block. Since the fanconi anemia pathway likewise plays an important role in regulating fragile sites (89), a direct role for activation of dormant origins in promoting genome stability under conditions of stress is one possible interpretation based on these finding.

Perspectives: the interplay between transcription regulation and origin activation

The activation of dormant origins and the increase in origin density during a perturbed cell cycle raises an important question concerning whether the modification of origin density is determined in G1 or occurs after S phase of the cell cycle begins. It was shown, for example, in the *X laevis* replication system that blocking the start of DNA replication does not result in the accumulation of a *trans* acting initiation factor and higher origin densities during S phase (39). This observation contrasts with the earlier observations in CHO cells that blocking the G1/S transition results in a time dependent increase in origin density. These observations can be reconciled if the factor that accumulates does so in G1 rather than S phase of the cell cycle (90, 91), since there is no G1 phase in the *X laevis* replication system. The accumulation of a licensing regulator or some other initiation factor during G1 in proportion to the time elapsed before S phase begins is consistent with the observations concerning the role of CMYC in globally regulating chromatin structure (92) and determining the replication program (65). Whether or not CMYC is one factor involved in regulating licensing and replicon size remains to be demonstrated, but its role as a regulator of G1 progression and gene transcription is consistent with such a proposal.

Alternatively, inhibiting initiation factors, such as Cdk2, might result in an enhanced loading of MCM proteins, and excessive licensing, during S phase (56, 57). Inhibiting Cdk2 results in checkpoint activation and excess MCM/CDC45 loading, but this observation is difficult to interpret given the apparently complete establishment of the replication program in G1. Recently, it was shown, however, that inhibiting CDC7 resulted in proportionally faster replication fork rates, and it would be interesting to see if the MCM/CDC45 loading is likewise enhanced under these conditions and thus accounts for the enhanced fork rates (93). The increased rate of fork movement might be explained, however, by other observations that the DNA helicase activity of the MCM4-6-7 complex is negatively regulated by CDK2 phosphorylation (94). Thus, reduced levels of CDC7 might likewise have a stimulating effect on fork movement, if it is associated with low CDK2 levels and correspondingly higher MCM/CDC45 activities.

Recently, it was proposed that replication fork rates and origin densities are co-regulated by a checkpoint mediated mechanism that coordinates RNR levels with origin densities (15). According to this proposal the checkpoint, or checkpoint related factors, imposes origin interference within a replication focus in order to establish a balance between replicon sizes and fork rates. Such a mechanism implies that RNR, Chk1 and pre-RCs co-localize to form a single regulatory unit, but the actual location of RNR in the cell remains uncertain in metazoans. A recent study, for example, suggests that RNR might translocate into the nucleus and allow dNTPs to initiate DNA synthesis under physiological conditions (95). Other studies have failed to detect a nuclear location of RNR in higher eukarotes. Nevertheless, dNTP pool sizes play in important role in regulating replication fork velocity and fidelity (96, 97).

The redundancy and flexibility in origin activation in eukaryotes suggests that replication fork movement determines when and where replication origins fire. If so, how the cell might coordinate fork movement with initiation sites is unclear, but emerging evidence suggests that

claspin/Mrc1 could play a crucial role due to its regulation of Chk1 activity and its effect on DNA synthesis (98). Inhibiting claspin, for example, reverses the UV-induced reduction of DNA synthesis, while degradation of claspin is associated with checkpoint attenuation (99, 100) and, possibly, increased origin firing. Accordingly, Claspin, which binds Chk1 upon checkpoint activation (101), might initially impose and subsequently relieve origin interference at immediately adjacent potential origins within a replicon cluster or replication focus. A central role for claspin in coordinating fork rates and origin densities awaits further investigation.

What is the relationship between replication origin usage and the regulation of gene expression? Embryonic cells are characterized by the absence of a G1 phase of the cell cycle and the absence of both gene transcription and checkpoint activation. The replication program in these systems is remarkably similar to what is observed in checkpoint compromised somatic cells. In embryos, origins are spaced on average every 5 to 15 Kb, which is similar to the spacing between potential origins in somatic cells (approximately 12 Kb). Moreover, fork rates in embryos are correspondingly slower than in somatic cells (0.6 Kb vs. 1.2 Kb per second), again reflecting the general inverse correlation between fork rates and origin densities (15).

These observations suggest that the developmental program that regulates gene transcription acts to coordinate transcription and replication during a normal cell cycle. Promoters, for example, have been increasingly associated with site specific replication origins (102, 103), and evidence is emerging that transcription factors play a direct role in determining site specific origin activation or repression. It has long been noted that MYC protein over production induces genomic instability (85, 104), while even transient overproduction can result extensive DNA damage (85, 104). This indicates that CMYC might act to coordinate the location and number of replication origins in G1 in accordance with the location and activity of gene promoters. In this manner, changes in the transcription pattern, or transcriptome, during cell differentiation, might dictate corresponding changes in the replication program in any particular tissue type. Future transcriptome based approaches to cellular differentiation and DNA replication promise to reveal how the hierarchy of origin efficiency and use is organized during development, and how the replicon is dynamically regulated in different cellular environments within the metazoan genome.

Based on the original suggestions made in the 1970s and 80s (25, 26, 43), the following model of dormant origin activation can be proposed: 1) DNA synthesis inhibitors, such as hydroxyurea, cause replication forks to stall; 2) dormant origins are induced up to 2X in a checkpoint independent manner as a function of the time of the replication block; 3) subsequent checkpoint activation due to the corresponding anomalous increase in fork density results in global origin inactivation; 4) prolonged checkpoint activation results in local Chk1 degradation (53); 5) heterochromatin spreads, and/or chromatin is remodelled possibly resulting in an embryonic state (26; see also105, 106 and 113); and 6) additional dormant origins fire as a component of replication re-start, resulting in an additional decrease in origin to origin distances (>2X).

Conversely, during development, CMYC levels are high in the early embryo, drop dramatically at the mid-blastula transition and increase in a tissue specific manner in somatic cells (107, 108). Corresponding changes in chromatin take place in a transcription dependent manner (109), thus restricting replication initiation to promoter regions. The apparently modular and hierarchical organization of potential replication origins maintains a balance between replication domain size and available replication factors, and thus results in a corresponding correlation between replicon size and DNA replication fork rates (15; Figure 4B). This organization plays a potentially important role in facilitating the cell's rapid exit from S phase when forks either stall or are blocked (110). Consequently, when fork movement is impeded the activation of dormant origins circumvents a delay in S phase progression due to stalled replication forks, and consequently protects the genome against a catastrophic accumulation of DNA damage.

Acknowledgements: The authors would like to thank Bianca Sclavi for helpful discussions.

Key words: replicon, dormant origin, DNA damage response, genome stability, checkpoint

Running title: Dormant origin activation and regulation of replicon size

Captions

Figure 1: Periodic spacing of initiation zones over an approximately 800 Kb region of 14q11.2 from human keratinocytes (adapted from from reference 31). Green: hybridization signals used to map initiation zones. Vertical white lines: initiation events that contributed to a zone during a given replication cycle. The average distance between zones (gap size) is approximately 25 Kb. Staggered bars correspond to replicons. Each initiation zone is either active (green) or inactive (red) during a given S phase (S1, S2, S3). Arrows represent bidirectional replication fork movement. The replicons are arbitrarily organized for purposes of illustration; origins corresponding to each replicon are randomly activated, and therefore will result in different patterns of replicon organization between successive S phases. The majority of the initiation zones map to intergenic regions in a manner unrelated to the level of gene expression. Bar = 100 Kb

Figure 2: Schematic of dormant, or backup, origins and replicon specification. ORC binding in G1 to a preferential origin (PO) is accompanied by a loading of excess MCM proteins at adjacent dormant origins (DO). During normal S phase (S), activation of a preferential origin results in local up-regulation of Chk1 in a surveillance mode (Boxes) and a subsequent block to dormant origin firing. When DNA damage occurs, ATR phosphorylates Chk1 (black dots) and activates the checkpoint. Prolonged stalling of forks results in the induction of Y family polymerases (pol β, κ), a concomitant down-regulation of Chk1 presumably by Skp1/Cul1/F-box (SCF), or a related ubiquitin mediated (grey dots) proteasomal complex, and the subsequent activation of dormant origins (HR).

Figure 3: Replication mediated gene amplification (adapted from Shimizu et al. 2003; see also 82). Replicating submicroscopic DNA molecules form when DNA breaks (lightning bolt) occur at or near a matrix attachment region (MAR). Gene amplification (arrows) occurs via rolling circle replication, and/or recombination with other homologous submicroscopic DNA molecules (see 111). Integration into double minute chromosomes is associated with re-integration into the chromosome (CEN: centromeric; TEL: telomeric orientations). At this stage, sequences will be amplified in a tandem orientation, and initiate the formation of homogenously staining regions (HSR) via a Break-Fusion-Bridge mechanism (BFB), which generates large inverted repeats.

Figure 4: Model of fragile site (FRA) stability during a normal S phase, and induction when replication forks are blocked. A and B: Dormant origins fire when replication through a replication slow zone (RSZ) occurs during a normal S phase. The activation of dormant origins in the absence of a fork inhibitor such as aphidicolin (AP) guarantees complete duplication of the region, and obviates chromosomal breakage at unreplicated DNA during G2/M. When forks are blocked by AP, dormant origins fire and replisome collision occurs, thus inducing the fragile site (DNA fragmentation). Arrows: direction of fork movement. Small boxes: Chk1; Black dot: phosphorylation; Rectangles: induced dormant origins. Red hache marks: DNA damage and fragmentation.

Figure 5: Replication focus model of dormant origin activation and replicon remodelling during development. A. Replicons are spaced every 5 to 15 Kb in embryos and roughly coincide with chromatin loop size (20 to 25 Kb). The 12 Kb replicon size suggests that features of the chromatin such as its persistence length impose a lower limit on replicon size (25, 26, 112). This results in a modular organization of the genome for replication; replicons are "popped out" in units of 12 Kb to form larger replicons. This is associated with a concomitant redistribution of replication factors (magenta ovals), and a balance between replicon size and replication fork rate within the actively replicating focus. Chromatin remodelling occurs when transcription is resumed. Origins active earlier in development become dormant (DO), in part because of checkpoint surveillance (blue boxes). At this stage initiation is increasingly restricted to intergenic regions and possible gene promoters (green boxes). Transcription factors such as MYC, JUN and E2F1-5 (red triangle) participate in the tissue specific coordination and activation of transcription and replication origins. B: Checkpoint mediated pathway of dormant origin activation. Differentiated, tissue-specific chromatin reverts to de-differentiated embryonic chromatin (see 26) after prolonged replication stress. When DNA damage occurs (lightning bolt), reactive oxygen species are produced (ROS) and Chk1 (blue boxes) is activated (black dots) during the first four hours of stress (4h). Dormant origins fire in a time dependent and checkpoint independent manner (passive firing). Origin firing up-regulates the checkpoint response as a consequence of the limited increase (~ 2X) in origin density, which signals the ubiquitin-proteasomal degradation of Chk1 (4 to 8h) and local checkpoint attenuation and replication re-start from dormant

origins as well as stalled forks. The spreading of heterochromatin is associated with a burst of origin activation and, eventually, Cdc45/MCM dependent apoptosis (see 72). Hence, there are two phases of dormant origin activation: 1) early stages of stress result in hyper-replication (DO activation) and upregulation of the DDR; and 2) later stages of prolonged stress result in checkpoint attenuation, massive DO activation and senescence/apoptosis.

References

1. Jacob, F., Brenner, S. & Cuzin, F. (1963) On the regulation of DNA replication in bacteria. *Cold Spring Harbor Symp. Quant. Biol.*, **28**, 329–348.

2. Skarstad K., Boye E, and Fanning E. (2003) Circles in the sand. *EMBO Rep.,* **4**, 661-5.

3. Aladjem M.I., and Fanning E. (2004) The replicon revisited: an old model learns new tricks in metazoan chromosomes. *EMBO Rep.,* **5**, 686-91.

4. Berezney R., Dharani D., Dubey D.D., and Huberman JA (2000) Heterogeneity of eukaryotic replicons, replicon clusters, and replication foci. *Chromosoma* **108,** 471-84.

5. Stinchcomb D.T., Struhl K., and Davis R.W. (1979) Isolation and characterisation of a yeast chromosomal replicator. *Nature* **282**, 39-43.

6. Kearsey S. (1983) Analysis of sequences conferring autonomous replication in baker's yeast. *EMBO J.* **2**, 1571-5

7. Bell S.P., and Stillman B. (1992) ATP-dependent recognition of eukaryotic origins of DNA replication by a multiprotein complex. *Nature* **357**, 128-34

8. Cvetic C., and Walter J.C. (2005) Eukaryotic origins of DNA replication: could you please be more specific? *Semin. Cell Dev. Biol.* **16**, 343-53.

9. Biamonti G., Paixão S., Montecucco A., Peverali F.A., Riva S., and Falaschi A. (2003) Is DNA sequence sufficient to specify DNA replication origins in metazoan cells? *Chromosome Res.* **11** 403-12

10. Huberman JA, Riggs AD. (1968) On the mechanism of DNA replication in mammalian chromosomes. *J. Mol. Biol.* **32**, 327-41

11. Edenberg H.J., and Huberman J.A.. (1975) Eukaryotic chromosome replication *Annu Rev Genet.* **9**, 245-84

12. Hand R. (1975) Regulation of DNA replication on subchromosomal units of mammalian cells. *J. Cell Biol.* **4**, 89-97

13. Hand R. (1978) Eucaryotic DNA: organization of the genome for replication. *Cell* **15** 317-25

14. Liapunova N.A. (1994) Organization of replication units and DNA replication in mammalian cells as studied by DNA fiber radioautography. *Int Rev Cytol.* **154**, 261-308.

15. Herrick J, Bensimon A. (1999) Single molecule analysis of DNA replication. Biochimie. **81**, 859-71.

16. Jackson D.A., and Pombo A. (1998) Replicon clusters are stable units of chromosome structure: evidence that nuclear organization contributes to the efficient activation and propagation of S phase in human cells. *J Cell Biol.* **140**, 1285-95

17. Herrick J., Stanislawski P., Hyrien O., and Bensimon A. (2000) Replication fork density increases during DNA synthesis in X. laevis egg extracts. *J Mol Biol.* **300,** 1133-42

19. Blow J.J., Gillespie P.J., Francis D., and Jackson D.A. (2001) Replication origins in Xenopus egg extract Are 5-15 kilobases apart and are activated in clusters that fire at different times. *J Cell Biol.* **152**, 15-25

18. Marheineke K., and Hyrien O. (2001) Aphidicolin triggers a block to replication origin firing in Xenopus egg extracts. *J. Biol. Chem.* **276**, 17092-100.

20. Takebayashi S.I., Manders E.M., Kimura H., Taguchi H., and Okumura K. (2001) Mapping sites where replication initiates in mammalian cells using DNA fibers. *Exp Cell Res.* 271(2):263-8.

21. Norio P., and Schildkraut C.L. (2001) Visualization of DNA replication on individual Epstein-Barr virus episomes, *Science*. **294**, 2361-4.

22. Patel P.K., Arcangioli B., Baker S.P., Bensimon A., and Rhind N. (2006) DNA replication origins fire stochastically in fission yeast. *Mol. Biol. Cell.* **17**, 308-16

23. Czajkowsky D.M., Liu J., Hamlin J.L., and Shao Z. (2007) DNA Combing Reveals Intrinsic Temporal Disorder in the Replication of Yeast Chromosome VI.
*J. Mol. Biol.*

24. Breier A.M., Chatterji S., and Cozzarelli N.R. (2004) Prediction of Saccharomyces cerevisiae replication origins. *Genome Biol.* **5**, R22

25. Taylor, J.H. (1977) Increase in DNA replication sites in cells held at the beginning of S phase. *Chromosoma.* **62**, 291–300

26. Taylor, J.H., and Hozier J.C. (1976) Evidence for a four micron replication unit in CHO cells. *Chromosoma* **57**, 341-50

27. Gilbert, D.M. (2007) Replication origin plasticity, Taylor-made: Inhibition vs. recruitment of origins under conditions of replication stress. *Chromosoma* **116,** 341–347

28. Anglana M., Apiou F., Bensimon A., and Debatisse M. (2003) Dynamics of DNA replication in mammalian somatic cells: nucleotide pool modulates origin choice and interorigin spacing. *Cell* **114**, 385-94

29. Alvino, G.M., Collingwood, D., Murphy J.M., Delrow J., Brewer B.J.,and Raghuraman M.K. (2007) Replication in hydroxyurea: it's a matter of time.
*Mol. Cell. Biol.* **27**, 6396-406

30. Norio P., Kosiyatrakul S., Yang Q., Guan Z., Brown N.M., Thomas S., Riblet R., and Schildkraut C.L. (2005) Progressive activation of DNA replication initiation in large domains of the immunoglobulin heavy chain locus during B cell development. Mol. Cell. **20**, 575-87

31. Lebofsky R., Heilig R., Sonnleitner M., Weissenbach J., a,d Bensimon A. (2006) DNA replication origin interference increases the spacing between initiation events in human cells. *Mol. Biol. Cell.* **17**, 5337-45

32. Edwards M.C., Tutter A.V., Cvetic C., Gilbert C.H., Prokhorova T.A., and Walter J.C. (2002) MCM2-7 complexes bind chromatin in a distributed pattern surrounding the origin recognition complex in Xenopus egg extracts. *J. Biol. Chem.* **277**, 33049-57

33. Hyrien O., Marheineke K., and Goldar A. (2003) Paradoxes of eukaryotic DNA replication: MCM proteins and the random completion problem. *Bioessays* **25**, 116-25.

34. Woodward A.M., Gohler T., Luciani M.G., Oehlmann M., Ge X., Gartner A., Jackson D.A., and Blow J.J. (2006) Excess Mcm2-7 license dormant origins of replication that can be used under conditions of replicative stress. *J. Cell. Biol.* **173**, 673-83

35. Ge X.Q., Jackson D.A., and Blow J.J. (2007) Dormant origins licensed by excess Mcm2-7 are required for human cells to survive replicative stress. *Genes Dev.* **21**, 3331-41

36. Teer J.K., Machida Y.J., Labit H., Novac O., Hyrien O., Marheineke K., Zannis-Hadjopoulos M., and Dutta A. (2006) Proliferating human cells hypomorphic for origin recognition complex 2 and pre-replicative complex formation have a defect in p53 activation and Cdk2 kinase activation.

*J. Biol. Chem.* **281**, 6253-60.

37. Miao H., Seiler J.A., and Burhans W.C. (2003) Regulation of cellular and SV40 virus origins of replication by Chk1-dependent intrinsic and UVC radiation-induced checkpoints, *J. Biol. Chem.* **278**, 4295-304.

38. Shechter D., Costanzo V., and Gautier J. (2004) ATR and ATM regulate the timing of DNA replication origin firing, *Nat. Cell Biol.* **6**, 648-55.

39. Marheineke K., and Hyrien O. (2004) Control of replication origin density and firing time in Xenopus egg extracts: role of a caffeine-sensitive, ATR-dependent checkpoint, *J. Biol. Chem.* **279**, 28071-81.

40. Syljuåsen R.G., Sørensen C.S., Hansen L.T., Fugger K., Lundin C., Johansson F., Helleday T., Sehested M., Lukas J., and Bartek J. (2005) Inhibition of human Chk1 causes increased initiation of DNA replication, phosphorylation of ATR targets, and DNA breakage, *Mol. Cell. Biol.* **25**, 3553-62.

41. Zhang Y.W., Otterness D.M., Chiang G.G., Xie W., Liu Y.C., Mercurio F., Abraham R.T. (2005) Genotoxic stress targets human Chk1 for degradation by the ubiquitin-proteasome pathway, *Mol Cell.* **19**, 607-18.

42. Mamely I., van Vugt M.A., Smits V.A., Semple J.I., Lemmens B., Perrakis A., Medema R.H., and Freire R. (2006) Polo-like kinase-1 controls proteasome-dependent degradation of Claspin during checkpoint recovery, *Curr. Biol.* **16**, 1950-5.

43. Painter, R.B. (1985) Inhibition and recovery of DNA synthesis in human cells after exposure to ultraviolet light. *Mutat. Res.* **145**, 63–69

44. Zannis-Hadjopoulos M., Taylor M.W., and Hand R. (1980) Inhibition of DNA chain elongation in a purine-auxotrophic mutant of Chinese hamster. *J. Cell Biol.* **85**, 777-85

45. MacDougall C.A., Byun T.S., Van C., Yee M.C., and Chimprich K.A. (2007) The structural determinants of checkpoint activation. *Genes Dev*. **21**, 898-903

46. Pillaire M.J., Betous R., Conti C., Czaplicki J., Pasero P., Bensimon A., Cazaux C., and Hoffmann J.S. (2007) Upregulation of error-prone DNA polymerases beta and kappa slows down fork progression without activating the replication checkpoint. *Cell Cycle*. **6**, 471-7

47. Bi X., Barkley L.R., Slater D.M., Tateishi S., Yamaizumi M., Ohmori H., and Vaziri C. (2006) Rad18 regulates DNA polymerase kappa and is required for recovery from S-phase checkpoint-mediated arrest.
*Mol. Cell. Biol.* **26**, 3527-40

48. Bi X., Slater D.M., Ohmori H., and Vaziri C. (2005) DNA polymerase kappa is specifically required for recovery from the benzo[a]pyrene-dihydrodiol epoxide (BPDE)-induced S-phase checkpoint.
*J. Biol. Chem.* **280**, 22343-55

49. Barkley L.R., Ohmori H., and Vaziri C. (2007) Integrating S-phase checkpoint signaling with trans-lesion synthesis of bulky DNA adducts. *Cell. Biochem. Biophys.* **47**, 392-408

50. Luke B., Versini G., Jaquenoud M., Zaidi I.W., Kurz T., Pintard L., Pasero P., and Peter M. (2006) The cullin Rtt101p promotes replication fork progression through damaged DNA and natural pause sites, *Curr. Biol.* **16**, 786-92.

51. Santocanale C., Sharma K., and Diffley JF. (1998) A Mec1- and Rad53-dependent checkpoint controls late-firing origins of DNA replication, *Nature* **395**, 615-8

52. Shirahige K., Hori Y., Shiraishi K., Yamashita M., Takahashi K., Obuse C., Tsurimoto T., and Yoshikawa H. (1998) Regulation of DNA-replication origins during cell-cycle progression, *Nature* **395**, 618-21.

53. Zhang Y.W., Otterness D.M., Chiang G.G., Xie W., Liu Y.C., Mercurio F., and Abraham R.T. (2005) Genotoxic stress targets human Chk1 for degradation by the ubiquitin-proteasome pathway.
*Mol. Cell* **19**, 607-18

54. Liu P., Barkley L.R., Day T., Bi X., Slater D.M., Alexandrow M.G., Nasheuer H.P., and Vaziri C. (2006) The Chk1-mediated S-phase checkpoint targets initiation factor Cdc45 via a Cdc25A/Cdk2-independent mechanism.
*J. Biol. Chem.* **281**, 30631-44.

55. Heffernan T.P., Unsal-Kaçmaz K., Heinloth A.N., Simpson D.A., Paules R.S., Sancar A., Cordeiro-Stone M., and Kaufmann W.K. (2007) Cdc7-Dbf4 and the human S checkpoint response to UVC.
*J. Biol. Chem.* **282**, 9458-68

56. Zhu Y., Alvarez C., Doll R., Kurata H., Schebye X.M., Parry D., and Lees E. (2005) Human CDK2 inhibition modifies the dynamics of chromatin-bound minichromosome maintenance complex and replication protein A.
*Cell Cycle*. **4** 1254-63.

57. Zhu Y., Alvarez C., Doll R., Kurata H., Schebye X.M., Parry D., and Lees E. (2004) Intra-S-phase checkpoint activation by direct CDK2 inhibition.
*Mol. Cell. Biol.* **24**, 6268-77

58. Hua X.H., Yan H., and Newport J. (1997) A role for Cdk2 kinase in negatively regulating DNA replication during S phase of the cell cycle.
*J. Cell Biol.* **137**, 183-92

59. Koppen A., Ait-Aissa R., Koster J., van Sluis P.G., Ora I., Caron H.N., Volckmann R., Versteeg R., and Valentijn L.J. (2007). Direct regulation of the minichromosome maintenance complex by MYCN in neuroblastoma.
*Eur. J. Cancer.* **43**, 2413-22.

60. Secombe J., Pierce S.B., and Eisenman R.N. (2004) Myc: a weapon of mass destruction.
*Cell* **117**, 153-6.

61. Berns K., Hijmans E.M., and Bernards R. Repression of c-Myc responsive genes in cycling cells causes G1 arrest through reduction of cyclin E/CDK2 kinase activity.
*Oncogene* **15**, 1347-56

62. Deb-Basu D., Karlsson A., Li Q., Dang C.V., and Felsher D.W. (2006) MYC can enforce cell cycle transit from G1 to S and G2 to S, but not mitotic cellular division, independent of p27-mediated inihibition of cyclin E/CDK2.
*Cell Cycle* **5**, 1348-55

63. Deb-Basu D., Aleem E., Kaldis P., and Felsher D.W. (2006) CDK2 is required by MYC to induce apoptosis.
*Cell Cycle* **5**, 1342-7.

64. Ray S., Atkuri K.R., Deb-Basu D., Adler A.S., Chang H.Y., Herzenberg L.A., and Felsher D.W. (2006) MYC can induce DNA breaks in vivo and in vitro independent of reactive oxygen species.
*Cancer Res*. **66**, 6598-605

65. Dominguez-Sola D., Ying C.Y., Grandori C., Ruggiero L., Chen B., Li M., Galloway D.A., Gu W., Gautier J., and Dalla-Favera R. (2007) Non-transcriptional control of DNA replication by c-Myc.
*Nature* **448**, 445-51

66. Lebofsky R., and Walter J.C. (2007) New Myc-anisms for DNA replication and tumorigenesis?
*Cancer Cell* **12**, 102-3.

67. Hook S.S., Lin J.J., and Dutta A. (2007) Mechanisms to control rereplication and implications for cancer.

*Curr. Opin. Cell Biol.* **19**, 663-71.

68. Sivaprasad U., Machida Y.J., and Dutta A. (2007) APC/C--the master controller of origin licensing?
*Cell Div.* **23**, 8

69. Di Micco R., Fumagalli M., Cicalese A., Piccinin S., Gasparini P., Luise C., Schurra C., Garre' M., Nuciforo P.G., Bensimon A., Maestro R., Pelicci P.G., and d'Adda di Fagagna F. (2006) Oncogene-induced senescence is a DNA damage response triggered by DNA hyper-replication.
*Nature* **444**, 638-42.

70. Campisi J., and d'Adda di Fagagna F. (2007) Cellular senescence: when bad things happen to good cells.
*Nat. Rev. Mol. Cell Biol.* **8**, 729-40.

71. Schulz T.J., Zarse K., Voigt A., Urban N., Birringer M., and Ristow M. (2007) Glucose restriction extends Caenorhabditis elegans life span by inducing mitochondrial respiration and increasing oxidative stress.
*Cell Metab.* **6**, 280-93

72. Rodriguez R., Gagou M.E., and Meuth M. (2008) Apoptosis induced by replication inhibitors in Chk1-depleted cells is dependent upon the helicase cofactor Cdc45.
*Cell Death Differ.* **15**, 889-98.

73. Cahill D.P., Kinzler K.W., Vogelstein B., and Lengauer C. (1999) Genetic instability and darwinian selection in tumours.
*Trends Cell Biol.* **9**, M57-60.

74. Draviam V.M., Xie S., and Sorger P.K. (2004) Chromosome segregation and genomic stability.
*Curr. Opin. Genet. Dev.* **14**, 120-5.

75. Varshavsky A. (1981) On the possibility of metabolic control of replicon "misfiring": relationship to emergence of malignant phenotypes in mammalian cell lineages.
*Proc. Natl. Acad. Sci. U. S. A.* **78**, 3673-7.

76. Schimke R.T., Sherwood S.W., Hill A.B., and Johnston R.N. (1986) Overreplication and recombination of DNA in higher eukaryotes: potential consequences and biological implications.
*Proc. Natl. Acad. Sci. U. S. A.* **83**, 2157-61

77. Hoy C.A., Rice G.C., Kovacs M., and Schimke R.T. (1987) Over-replication of DNA in S phase Chinese hamster ovary cells after DNA synthesis inhibition.
*J. Biol. Chem.* **262**, 11927-34

78. Johnston R.N., Feder J., Hill A.B., Sherwood S.W., and Schimke R.T. (1986) Transient inhibition of DNA synthesis results in increased dihydrofolate reductase synthesis and subsequent increased DNA content per cell.
*Mol. Cell. Biol.* **6**, 3373-81

79. Albertson D.G. (2006) Gene amplification in cancer. *Trends Genet.* **22**, 447-55

80. Hahn P., Kapp L.N., Morgan W.F., and Painter R.B. (1986) Chromosomal changes without DNA overproduction in hydroxyurea-treated mammalian cells: implications for gene amplification.
*Cancer Res.* **46**, 4607-12.

81. Myllykangas S., and Knuutila S. (2006) Manifestation, mechanisms and mysteries of gene amplifications.
*Cancer Lett.* **232**, 79-89.

82. Von Hoff D.D., Needham-VanDevanter D.R., Yucel J., Windle B.E., and Wahl G.M. (1988) Amplified human MYC oncogenes localized to replicating submicroscopic circular DNA molecules.
*Proc. Natl. Acad. Sci. U. S. A.* **85**, 4804-8

83. Shimizu N., Hashizume T., Shingaki K., and Kawamoto J.K. (2003) Amplification of plasmids containing a mammalian replication initiation region is mediated by controllable conflict between replication and transcription.
*Cancer Res.* **63**, 5281-90

84. Davidson I.F., Li A., and Blow J.J. (2006) Deregulated replication licensing causes DNA fragmentation consistent with head-to-tail fork collision.
*Mol. Cell.* **24**, 433-43

85. Felsher D.W., and Bishop J.M. (1999) Transient excess of MYC activity can elicit genomic instability and tumorigenesis.
*Proc. Natl. Acad. Sci. U. S. A.* **96**, 3940-4

86. Cha R.S., and Kleckner N. (2002) ATR homolog Mec1 promotes fork progression, thus averting breaks in replication slow zones.
*Science* **297**, 602-6

87. . Durkin S.G., Arlt M.F., Howlett N.G., and Glover T.W. (2006) Depletion of CHK1, but not CHK2, induces chromosomal instability and breaks at common fragile sites.
*Oncogene* **25**, 4381-8.

88. Grossi S. Decaillet C., and Constantinou C. (2007) Role of Fanconi anemia pathway in recovery from a hydroxyurea replication block. Submitted to *DNA Repair*

89. Howlett N.G., Taniguchi T., Durkin S.G., D'Andrea A.D., and Glover T.W. (2005) The Fanconi anemia pathway is required for the DNA replication stress response and for the regulation of common fragile site stability.
*Hum. Mol. Genet.* **14**, 693-701.

90. Obaya A.J., Kotenko I., and Cole M.D., and Sedivy J.M. (2002) The proto-oncogene c-myc acts through the cyclin-dependent kinase (Cdk) inhibitor p27(Kip1) to facilitate the activation of Cdk4/6 and early G(1) phase progression.
*J. Biol. Chem.* **277**, 31263-9.

91. Santoni-Rugiu E., Falck J., Mailand N., Bartek J., and Lukas J. (2000) Involvement of Myc activity in a G(1)/S-promoting mechanism parallel to the pRb/E2F pathway. *Mol. Cell. Biol.* **20**, 3497-509

92. Knoepfler P.S., Zhang X.Y., Cheng P.F., Gafken P.R., McMahon S.B., Eisenman R.N. (2006) Myc influences global chromatin structure. *EMBO J.* **25**, 2723-34.

93. Montagnoli A., Menichincheri M., Tibolla M., Tenca P., Rainoldi S., Brotherton D., Valsasina B., Croci V., Albanese C., Patton V., Alzani R., Ciavolella A., Sola F., Molinari A., Volpi D., Bensimon A., Vanotti E., and Santocanale C. (2008) submitted *Nat. Chem. biol.*

94. Ishimi Y., Komamura-Kohno Y., Kwon H.J., Yamada K., and Nakanishi M. (2003) Identification of MCM4 as a target of the DNA replication block checkpoint system. *J. Biol. Chem.* **278**, 24644-50.

95. Liu X., Zhou B., Xue L., Shih J., Tye K., Qi C., and Yen Y. (2005) The ribonucleotide reductase subunit M2B subcellular localization and functional importance for DNA replication in physiological growth of KB cells. *Biochem. Pharmacol.* **70**, 1288-97.

96. Malinsky J., Koberna K., Stanek D., Masata M., Votruba I., and Raska I. (2001) The supply of exogenous deoxyribonucleotides accelerates the speed of the replication fork in early S-phase. *J Cell Sci.* **114**, 747-50

97. Mathews C.K. (2006) DNA precursor metabolism and genomic stability. *FASEB J.* **20**, 1300-14.

98. Chini C.C., and Chen J. (2003) Human claspin is required for replication checkpoint control. *J. Biol. Chem.* **278**, 30057-62.

99. Gewurz B.E., and Harper J.W. (2006) DNA-damage control: Claspin destruction turns off the checkpoint.

*Curr. Biol.* **16**, :R932-4

100. Bennett L.N., and Clarke P.R. (2006) Regulation of Claspin degradation by the ubiquitin-proteosome pathway during the cell cycle and in response to ATR-dependent checkpoint activation.
*FEBS Lett.* **580**, 4176-81.

101. Chini C.C., Wood J., and Chen J. (2006) Chk1 is required to maintain claspin stability.
*Oncogene* **25**, 4165-71

102. Huvet M., Nicolay S., Touchon M., Audit B., d'Aubenton-Carafa Y., Arneodo A., and Thermes C. (2007) Human gene organization driven by the coordination of replication and transcription.
*Genome Res.* **17**, 1278-85.

103. Danis E., Brodolin K., Menut S., Maiorano D., Girard-Reydet C., and Méchali M. (2004) Specification of a DNA replication origin by a transcription complex.
*Nat. Cell. Biol.* **6**, 721-30.

104. Karlsson A., Deb-Basu D., Cherry A., Turner S., Ford J., and Felsher D.W. (2003) Defective double-strand DNA break repair and chromosomal translocations by MYC overexpression.
*Proc. Natl. Acad. Sci. U. S. A.* **100**, 9974-9.

105. Singh G., and Klar A.J. (2008) Mutations in deoxyribonucleotide biosynthesis pathway cause spreading of silencing across heterochromatic barriers at the mating-type region of the fission yeast.
*Yeast* **25**, 117-28

106. Fahrner J.A., and Baylin S.B. (2003) Heterochromatin: stable and unstable invasions at home and abroad.
*Genes Dev.* **17**, 1805-12.

107. Lemaitre J.M., Bocquet S., Buckle R., and Mechali M. (1995) Selective and rapid nuclear translocation of a c-Myc-containing complex after fertilization of Xenopus laevis eggs.
*Mol. Cell. Biol.* **15**, 5054-62

108. King M.W., Roberts J.M., and Eisenman R.N. (1986) Expression of the c-myc proto-oncogene during development of Xenopus laevis.
*Mol. Cell. Biol.* **6**, 4499-508

109. Knoepfler P.S., Zhang X.Y., Cheng P.F., Gafken P.R., McMahon S.B., and Eisenman R.N. (2006) Myc influences global chromatin structure.
*EMBO J.* **25**, 2723-34.

110. Herrick and Bensimon Introduction to Molecular Combing: Genomics, DNA replication and Cancer Humana Press

111. Watanabe T., and Horiuchi T. (2005) A novel gene amplification system in yeast based on double rolling-circle replication.
*EMBO J.* **24**, 190-8.

112. Jun S, Herrick J, Bensimon A, Bechhoefer J. (2004) Persistence length of chromatin determines origin spacing in Xenopus early-embryo DNA replication: quantitative comparisons between theory and experiment.
*Cell Cycle.* **3**, 223-9.

113. Courbet S, Gay S, Arnoult N, Wronka G, Anglana M, Brison O, Debatisse M.(2008) Replication fork movement set chromatin loop size and origin choice in mammalian cells.
*Nature.* **455**:557-60

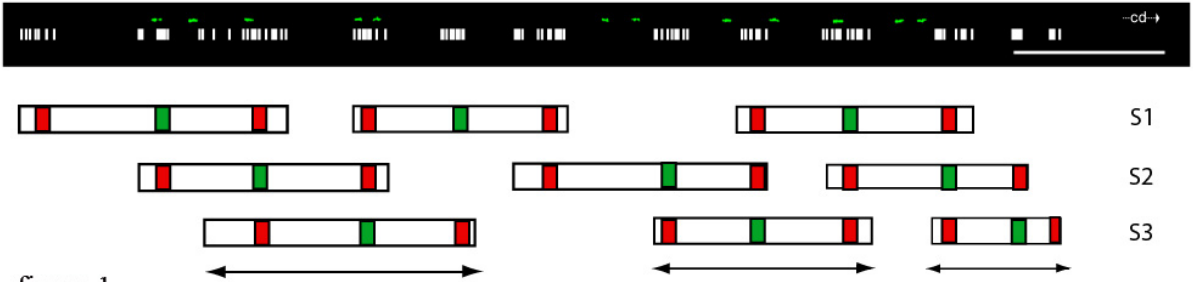

figure 1

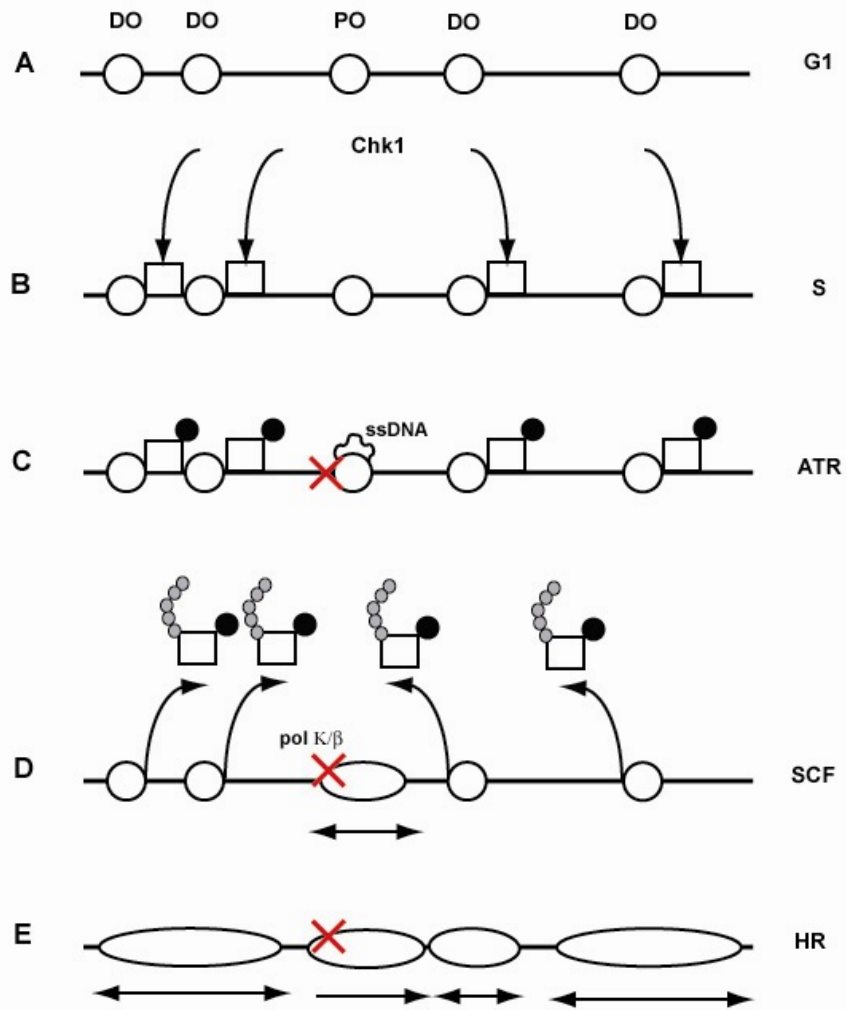

figure 2

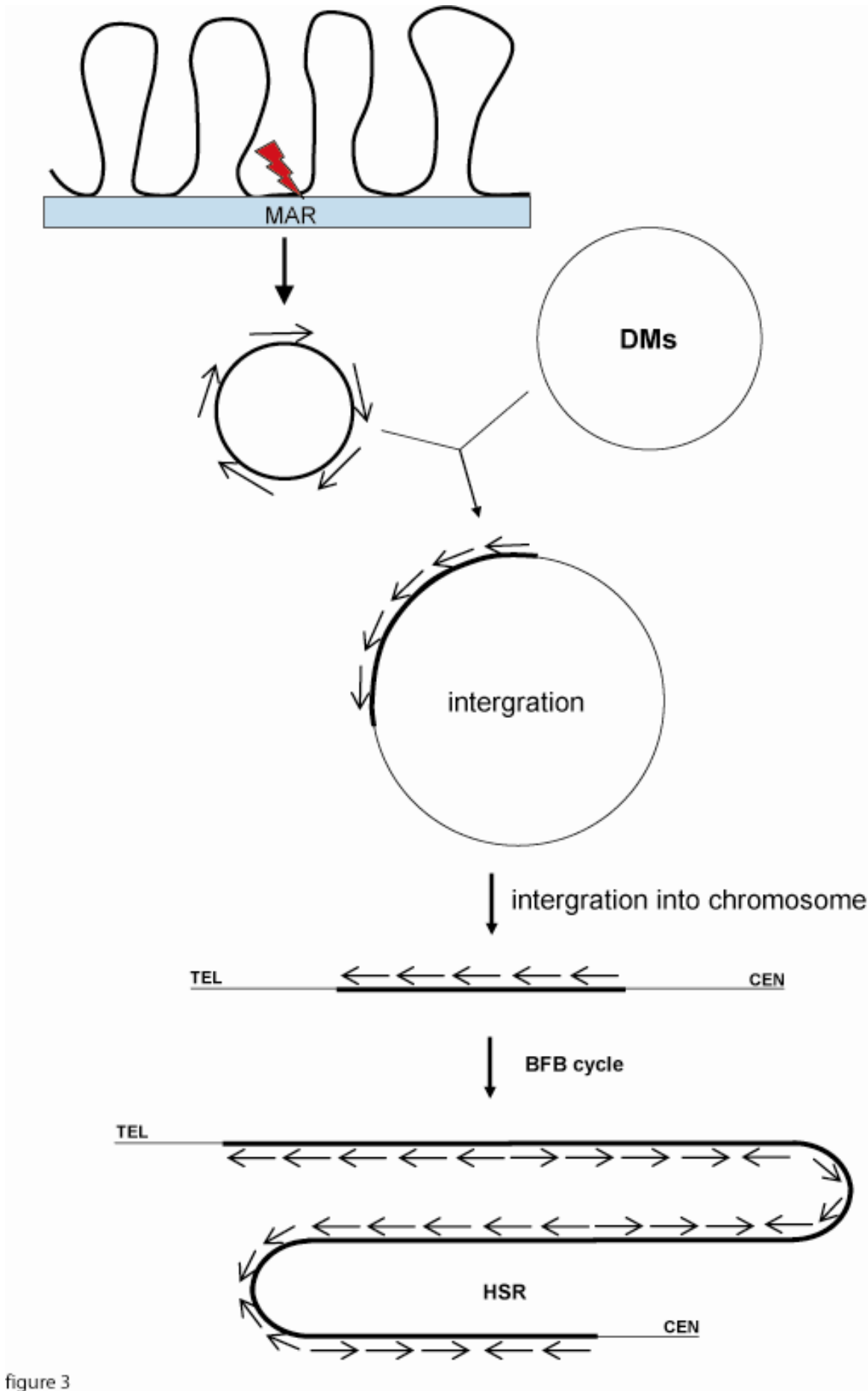

figure 3

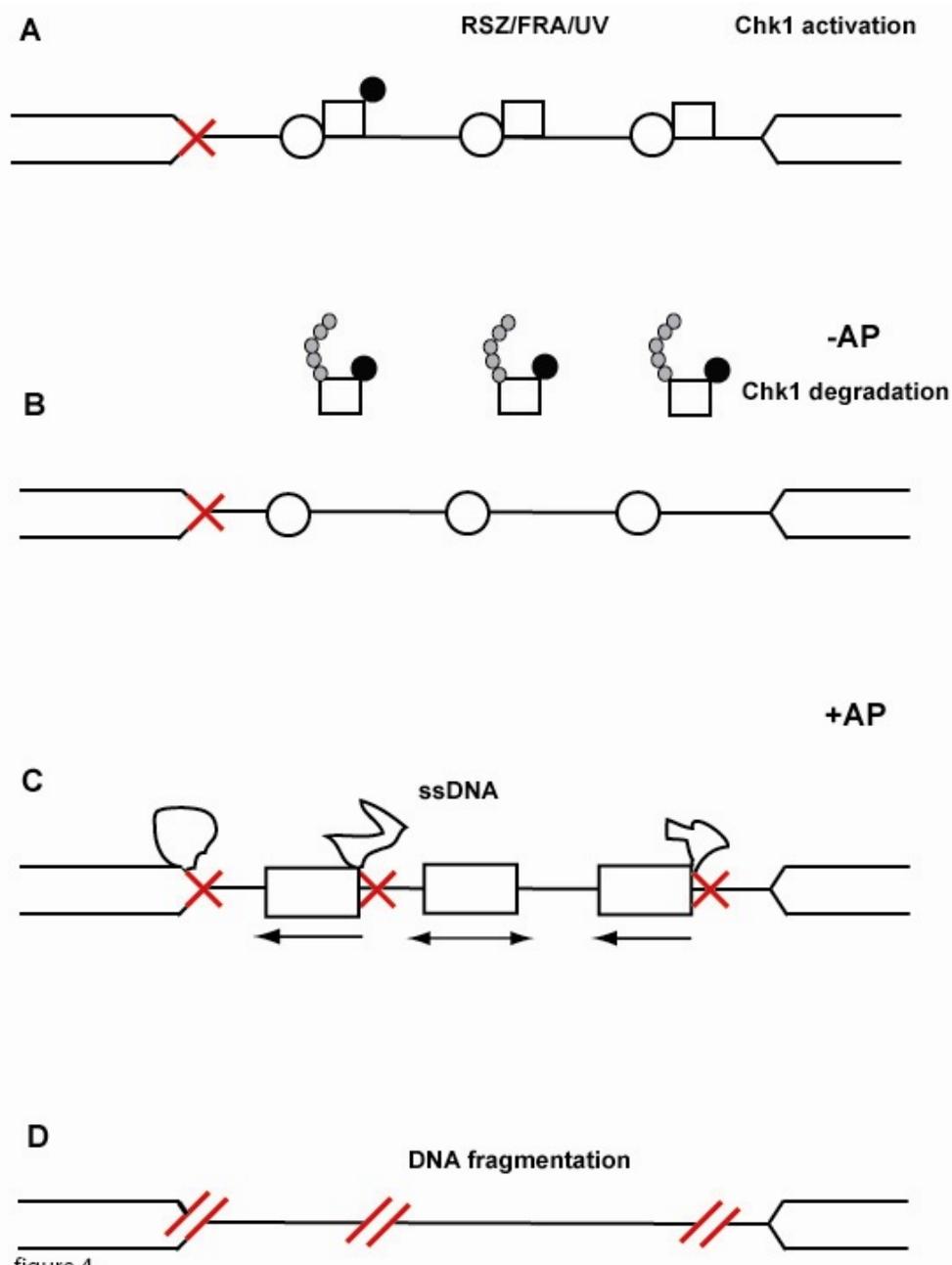

figure 4

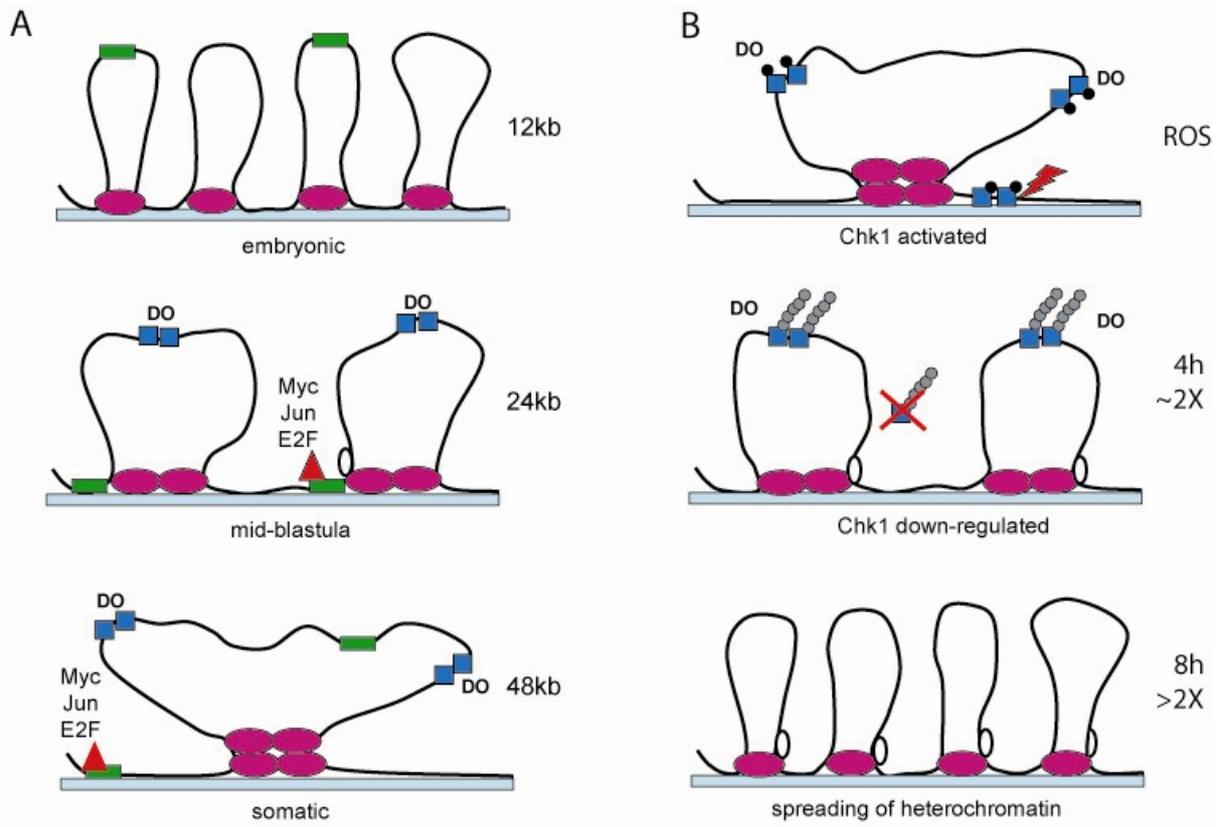
figure 5